\documentclass{ws-procs9x6}

\begin{document}

\title{Collectivity, chaos, and computers}

\author{Calvin W.~Johnson}

\address{Department of Physics, San Diego State University\\
5500 Campanile Drive\\
San Diego, CA 92182-1233, USA\\
E-mail: johnson@physics.sdsu.edu}

\maketitle
\abstracts{
Two important pieces of nuclear structure are many-body
collective deformations and single-particle spin-orbit splitting.
The former can be well-described microscopically by simple SU(3)
irreps,
but the latter mixes SU(3) irreps, which presents a challenge
for large-scale, \textit{ab initio} calculations on fast
modern computers. Nonetheless, SU(3)-like
phenomenology remains even in the face of strong mixing. The
robustness of band structure is reminiscent of robust,
pairing collectivity that arises from random two-body
interactions.
}

\section{Nuclear epistemology}

The goal of nuclear structure theory is to understand
experimental spectra, but what do we mean by \textit{understand}?

There are two routes to understanding. One is the brute force,
extreme reductionism of \textit{ab initio} computations\cite{abinitio}:
one starts from bare nucleon-nucleon scattering data and eventually computes
many-body binding energies, spectra, transition rates, etc.
This is of course very appealing to physicists, and I believe
such \textit{ab initio} calculations are among the most important
nuclear structure results of the past decade, but it has the
very obvious danger of getting lost in the numerical details.
Furthermore, brute-force calculations are limited by
available computing power.

The alternate is back-of-the-envelope reasoning: simple,
primarily analytic models that are easy to intuit.  Algebraic models
are prime examples\cite{simple,Rowe1996}.  The danger here is that the simple picture
may not accurately represent the microscopic physics.

Ideally, one would like to combine these two: perform microscopically
detailed \textit{ab initio} calculations built upon basis state
constructed from algebraic models. Such a hybrid approach
would certainly allow one greater insight into the microscopic
calculation, and would, one hopes, be much more efficient.

More specifically, I wish to address the possibility of using SU(3)
irreps
as a best basis for large-scale microscopic calculations.  The
biggest obstacle is spin-orbit
splitting that arises from the nuclear mean field.

\section{A brief guide to nuclear structure}

Nuclear structure is driven by several competing degrees of
freedom. First, the nucleus has a mean field, which allowed
Haxel, Jensen, and Suess, and
Mayer\cite{mayer49} to propose the non-interacting shell model. One of the
primary features of the nuclear mean field is a strong spin-orbit
splitting.  Spin-orbit splitting arises naturally from a non-relativistic
reduction of the Dirac equation; as the nucleus
is more relativistic than the atom, it is
understandable that spin-orbit splitting is very small in atomic
physics but is a large feature in nuclear physics. (In fact
it becomes so large that it gives rise to \textit{pseudospin};
see ref.~\cite{pseudospin,pseudospin_rel} and the contributions by Ginocchio
and van Isacker in this volume.) Eventually
from the noninteracting shell model developed the interacting
shell model, where one chooses a finite set of fermion shell model
states and diagonalizes a Hamiltonian in that space.

The starting point of the shell model is the independent particle
assumption: the component protons and neutrons interact primarily
with the mean field. This is a simplification: there are
correlations between the nucleons, and most important are the
collective correlations.  One well known form of collectivity is
pairing, whereby fermions of opposite (angular) momentum couple to
zero.  This is a pervasive feature of cold, dense fermion systems,
but it play only a peripheral role in this paper. Through the
seniority model\cite{simple} one can understand pairing in terms
of microscopic fermion states.

Instead the collectivity that I will pay most attention to, and the one
which has been the primary focus  of Jerry Draayer's work,
is quadrupole deformation. One can have both quadrupole vibrations
and ``static'' quadrupole deformations that lead to rotational bands.
 Quadrupole deformations arise naturally
out of the semiclassical liquid drop model of the nucleus, and
can be treated more formally in the Bohr-Mottelson model and its
generalization to geometric-collective models\cite{BM}.

One of the great breakthroughs in nuclear stucture physics was
discovering how to connect collective motion to the underlying
fermion microphysics. Elliot's
SU(3) model\cite{Ell58} and its successors showed how one could map
rotational motion easily onto the fermion shell model. Furthermore,
as Rowe\cite{Rowe1996} has emphasized, SU(3) maps also onto the Bohr-Mottelson
and similar models, thus providing a critical bridge between macroscopic
and microscopic pictures.  SU(3), at least as phenomenology,
describes beautifully many features of nuclear spectra. But how well
do microscopic SU(3) wavefunctions match `realistic' microscopic
wavefunctions? That is a question I will return to.


One of the most powerful tools for nuclear structure is the
spherical interacting shell model. Here one starts by assuming
a spherically symmetric mean field, so that all single-particle
states have good $j$. The model space is partitioned into subspaces
by single-particle configurations: one subspace
might be, for example, all states with the configuration
$(0d_{5/2})^3(1s_{1/2})^1 (0d_{3/2})^2$. In fact, because the
Hamiltonian is rotationally invariant, one can restrict to
the states with a fixed total $M$ (that is, $J_z$) and hence
programs that work in this basis are often referred to as $M$-scheme codes.
The many-body Hamiltonian matrix elements are then computed
in this basis.

Because the total angular momentum operator $J^2$ does not
connect across configurations, it is easy to construct a many-body
model space for which angular momentum is a good quantum number.
Futhermore, spin-orbit splitting can be treated nearly trivially in such
model spaces. What cannot be treated easily is deformation:
deformation mixes many configurations, and typically one needs to
add effective charges to get correct magnitudes for E2 transitions, etc.

Despite this important drawback,
$M$-scheme and related codes are very popular today. Some of the older
codes, such as the Glasgow code\cite{Whi77} or OXBASH\cite{oxbash}, store the many-body
Hamiltonian on disk.  This works for a basis size of up to about
half a million basis states. Beyond that, more recent shell-model
codes such as ANTOINE\cite{antoine} or REDSTICK\cite{redstick} recompute the Hamiltonian
many-body matrix elements on-the-fly.  With hard work and clever
coding, this can be very efficient.


The interacting shell model in a spherical basis
 is not the only possible approach.
One of Jerry Draayer's great achievements has been to construct,
with a series of collaborators, SU(3) shell model
codes\cite{su3me,su3shell,su3shell0,Bah95,Esc98}.
At the heart of these codes are Slater determinants
in  a \textit{cylindrical} rather than a spherical
single-particle basis. This allows one rather easily to get
fermion representations of SU(3).

A useful generalization of the SU(3) model is the symplectic
model \cite{symplectic}, which unifies quadrupole operators with
center-of-mass motion. This allows one to treat multi-$\hbar\Omega$
shell spaces and to project out exactly  spurious center-of-mass motion.
Futhermore one can generalize the SU(3) technology to symplectic
calculations\cite{symplectic_shell}.

Other properties, however, are not as easy. Projection
of good angular momentum is not as straightforward
as for spherical shell-model configurations.  (This can
probably be made more efficient.)

Unfortunately one remaining feature of nuclear structure remains
a potential obstacle:
spin-orbit splitting from the mean field, which I take up in the next
section.

\section{Mixed or pure SU(3)--that is the question}

 A number of well-known phenomenological
interactions mix SU(3) irreps. One is pairing\cite{Bah95}.  More germane to
my discussion is single-particle spin-orbit splitting which arises
from the nuclear mean-field\cite{Esc98,Gue00}. The bottom line:
in calculations in the $sd$- and lower $pf$-shells one finds that
single-particle
spin-orbit splitting is by far the most important source of mixing of SU(3) irreps.
If one eliminates spin-orbit splitting, then mixing of SU(3) irreps
is enormously reduced\cite{Gue00,Joh02}.

We investigated the role of spin-orbit splitting as
follows\cite{Gue00}. First, we took `realistic' interactions:
Wildenthal's USD interaction in the $sd$-shell \cite{wildenthal}
 and the monopole-modified KB3 interaction
in the $pf$-shell \cite{KB3}. These interactions started life as exact $G$-matrix
effective interactions reduced from nucleon-nucleon forces,
with some empirical adjustments fit to hundreds of levels and decays.
These interactions are by no means schematic and
were derived blindly with respect to SU(3).

We computed for various nuclides the `exact' wavefunctions for these complicated,
messy interactions. Using a Lanczos moment method, similar to that developed to
compute Gamow-Teller strength distributions\cite{Cau90}, we were able to compute the
distribution of the exact wavefunction onto SU(3) irreps, without having to
compute all the SU(3) eigenstates.

In the $sd$-shell and particularly in the $pf$-shell we found the wavefunctions
to be fragmented over many SU(3) irreps.  When we simply eliminated the spin-orbit
splitting, and nothing else, then the fragmentation was enormously reduced, even
for $pf$-shell nuclides.
One can fairly interpret
the fragmentation of SU(3) irreps due to realistic spin-orbit
splitting to mean that wavefunctions of pure or nearly pure SU(3) irreps are not
very realistic on a microscopic level.

Despite this fragmentation, the energy of deformation is larger
than spin-orbit splitting.  If one considers, for
example, $^{24}$Mg in the $sd$-shell, the leading (8,4) irrep of
SU(3) outperforms the simplest spherical configuration
$(d_{5/2})^8$, in terms of binding energy, B(E2) values,
etc.\cite{Gue02}.

While that is impressive, another calculation using a single
Hartree-Fock state projected onto good angular momentum, showed
that for a number of $sd$- and lower $pf$-shell nuclides the
projected HF state outperforms a single SU(3) irrep\cite{Joh02}.
Again the difference is driven almost entirely by the
single-particle spin-orbit splitting: if one removes spin-orbit
splitting, the difference between a projected HF state and the
leading irrep is small. 

So clearly one needs \textit{both} deformation
\textit{and} spin-orbit splitting. Any route that neglects one 
over the other has to work hard to catch up.  But which 
route?  A shell-modeler faces a large number of choices for basis
states: spherical shell model
configuations; SU(3) irreps; configurations built upon deformed
Hartree-Fock\cite{deformshell}; or a 'mixed-mode'
basis\cite{Gue02} combining two or more of these. Which is best, 
and how can we tell which is `best'?

As discussed above, an SU(3) basis can be very illuminating in
terms of the physics. For very-large-scale calculations, however, one
must be concerned with computational effciency. For example, in
full $0\hbar\Omega$ shell-model calculations, codes working in a
spherical basis are still much more efficient than an SU(3) basis;
the former can compute $^{24}$Mg roughly ten times faster than the
latter. Let me emphasize that is for the \textit{full} space
including \textit{all} configurations; the motivation of using an
SU(3) basis is the belief that one can truncate drastically to a
smaller and more efficient basis and still get a very good
description of the spectrum and wavefunctions.

Unfortunately SU(3) isn't always as efficient as one would hope, due to mixing of
irreps due to spin-orbit splitting (much as quadrupole deformation strongly
mixes configurations in the spherical shell model).   In a hybrid approach,
Gueorgueiv et al\cite{Gue02} showed that an oblique--that is, nonorthogonal--basis consisting of a
few SU(3) irreps and a few spherical configurations could work very well, requiring only
a few states: the SU(3) irreps encoded deformation and the spherical configurations
encoded spin-orbit splitting.

While this sounds marvelous,
the problem with such a statement is that those few states are not easy to represent
in the computer. Therefore, in order to truly diagnose how \textit{efficient} a
basis is, I make the following observation and proposal.  Most modern, large-scale,
interacting shell-model codes use Slater determinants as the fundamental
internal representation. The Slater determinant is in a single-particle basis:
spherical, cylindrical, or other such as Hartree-Fock.  We can discuss
usefully the computational efficiency in terms of \textit{the number of
Slater determinants needed to represent a state or to project out a good
quantum number}.

(Let me also note that there is a significant difference between a computationally
efficient basis and one that illuminates the physics. A handful of states, while
computationally inefficient, can still shed significant light upon the general structure
of the state.  But it is also important,  when proposing a basis,
to distinguish between computational efficiency and ``physics efficiency.'')

One can project out states of good angular momentum in a spherical
shell-model space relatively efficiently, requiring only a few
tens or a few hundred Slater determinants, and all within a single
configuration. Spin-orbit splitting is trivial; and if one crosses
major harmonic-oscillator shells, it is possible to project out
spurious center-of-mass motion as well, as long as one includes
the right set of configurations. On the other hand, building in
deformation requires many configurations.

In the cylindrical shell model, one can get the leading SU(3)
irreps, and thus deformation, easily with just a few Slater
determinants. Furthermore, if one has multi-shell calculations and
uses the symplectic extension, it is possible to project out
spurious center-of-masss motion. On the other hand, as presently
written in the SU(3) shell-model codes, projection of good angular
momentum is not very efficient, requiring several hundred or even
more than a thousand Slater determinants. (Indeed, this is why,
for exactly the same model space, spherical shell-model codes are
faster than SU(3) codes; the latter can probably be sped up.)
Finally, spin-orbit splitting can only be handled by mixing many
irreps.

What about Hartree-Fock based states?  They include deformation and
spin-orbit splitting, and the amount of effort needed to project out
good angular momentum is roughly comparable to SU(3) irreps. Unfortunately,
I believe that \textit{consistent} projection of spurious center-of-mass motion could be
problematic (for reasons I do not have space to discuss here); and this is
critical for large-basis calculations.

Ideally one would like to combine \textit{ab initio} calculations with the lessons
learned from algebraic models. The latter describe deformation, and in the
symplectic extension can project out spurious center-of-mass motion, but are strongly
mixed by spin-orbit splitting. I think
some generalized approach is needed, such as the ``optimal basis states''\cite{Car02}
which combine symplectic states with generator coordinate methods.  Such a proposal
seems very appealing, but needs further study.

\section{Persistence of collectivity}

Despite mixing of SU(3) irreps by spin-orbit splitting, the
resulting spectra --not only energy levels but also B(E2)
ratios--can be described very well \textit{phenomenologically} by
SU(3).  That is to say, if one numerically solves a Hamiltonian of
the form $Q\cdot Q+ l\cdot s$, the resulting wavefunctions will
strongly mix SU(3) irreps, but one can fit the energy levels and
B(E2)s to analytic SU(3) predictions. In other words, the mixing,
while strong, appears to be coherent. Chairul Bahri (Draayer's
former student) and David Rowe term  this `quasi-dynamical
symmetry'\cite{Bah00} and relate it to `adiabatic decoupling of
colective motion along the lines of the Born-Oppenheimer
approximation.' They considered the symplectic shell model with
the Davidson interaction, and found when the wavefunctions are
fragmented over many SU(3) irreps, the spectra still look
remarkably like SU(3) rotors. So even when microscopic
wavefunctions are not good SU(3) states, the spectral properties
still \textit{look} like SU(3).

There are several lessons to take away from
this.  First, that SU(3) irreps are not intrinsically very good microscopic
wavefunctions--and spin-orbit splitting is mainly responsible. Second,
and paradoxically, SU(3)-like behavior is very robust, suggesting that
one might be able to coherently mix SU(3) wavefunction to get both the
microscopic description and the phenomenology correct.  This, I believe, is
\textit{critical} for application to large-scale, \textit{ab inito}
shell-model calculations.

The final lesson is that collective behavior does not appear very sensitive
to the details of the Hamiltonian.  This leads me to my next topic.

\subsection{Collectivity and random interactions}

The above results argue that collective behavior is robust even when one adds
`messy' or `noisy' pieces to an algebraic Hamiltonian. One can take
this to the extreme and ask: if one leaves off the algebraic Hamiltonian
altogether and just have a `noisy' or \textit{random} Hamiltonian, does
any collective behavior remain?

Surprisingly, the answer is yes.  If one has a random two-body
interaction in a fermion shell model, one sees robust signatures
of pairing collectivity\cite{Joh98}. If one has a random two-body
interaction in the interacting boson model, one sees robust
rotational and vibrational band structures\cite{Bij00}. (See also
Roelof Bijker's contribution.)  To date, however, no one has found
a convincing random ensemble for the fermion shell model that
gives rise to robust band structure, despite some
proposals\cite{dtbre}.

There have been a large number of papers written analyzing and purporting to
explain the pairing-like behavior in fermion models, but to my mind none
of them are terribly convincing. We still need a general theory of
how collective behavior can arise generically can arise, a more general version
of Bahri and Rowe's quasi-dynamical symmetry.  The situation reminds me of
quantum chaos: the quantum wavefunctions of classically
chaotic systems display `scars' of classical periodic, but unstable, orbits.

\section{Conclusion}

I lay out two challenges for the intersection of algebraic models with
large-basis shell-model diagonalization:

First, one must account for both deformation and spin-orbit splitting. 
If SU(3)-symplectic wavefunctions are to be an efficient
computational basis for large-scale \textit{ab initio}
calculations, we must generalize further to account for 
spin-orbit mixing \textit{a priori}. 

Second, we need to investigate further how collective
behavior arises generically and how it remains
robust even in the presence of messy interactions. A good 
explanation 
might help answer the first challenge. 

\section*{Acknowledgements}

It was a pleasure and an honor to be invited to give this overview talk
at the celebration of
Jerry Draayer's 60th birthday. The bibliography below gives just a taste of
his many contributions to nuclear structure theory.

The work described herein was funded by the U.S.~Department of Energy and the
National Science Foundation.

\end{document}